\documentclass[conference,final,10pt]{IEEEtran}
\usepackage[letterpaper, left=0.67in, right=0.67in, bottom=0.92in, top=0.76in]{geometry}
\usepackage[utf8]{inputenc}
\usepackage{caption,subcaption}
\usepackage{graphicx}
\usepackage{amsmath,amssymb,amsthm,mathtools}
\usepackage{color}
\usepackage{algorithm}
\usepackage{algpseudocode}
\usepackage{upgreek}
\usepackage{ragged2e}
\usepackage{psfrag}
\usepackage{stfloats}
\usepackage{float}
\usepackage{bm}
\usepackage{cite}

\usepackage[acronym,shortcuts]{glossaries}

\newacronym{5G}{5G}{fifth generation}
\newacronym{FD}{FD}{Full-Duplex}
\newacronym{MIMO}{MIMO}{Multiple-Input Multiple-Output}
\newacronym{SISO}{SISO}{Single-Input Single-Output}
\newacronym{HD}{HD}{Half-Duplex}
\newacronym{RF}{RF}{radio frequency}
\newacronym{IoT}{IoT}{Internet of Things}
\newacronym{ITS}{ITS}{Intelligent Transportation Systems}
\newacronym{DoF}{DoF}{Degrees of Freedom}
\newacronym{SI}{SI}{Self-Interference}
\newacronym{SIC}{SIC}{self-interference cancellation}
\newacronym{GCS}{GCS}{geometric constellation shaping}
\newacronym{PCS}{PCS}{probabilistic constellation shaping}
\newacronym{CSI}{CSI}{Channel State Information}
\newacronym{MB}{MB}{Maxwell-Boltzmann}
\newacronym{AWGN}{AWGN}{Additive White Gaussian Noise}
\newacronym{NL}{NL}{nonlinear}
\newacronym{MI}{MI}{mutual information}
\newacronym{PDF}{PDF}{probability density function}
\newacronym{BER}{BER}{bit error rate}
\newacronym{CGR}{CGR}{Complex Gaussian Ratio}
\newacronym{D2D}{D2D}{device-to-devide}
\newacronym{MMSE}{MMSE}{minimum mean square error}
\newacronym{SINR}{SINR}{Signal-to-Interference-plus-noise Ratio}
\newacronym{ZF}{ZF}{Zero-Forcing}
\newacronym{MRT}{MRT}{Maximum Ratio Transmission}
\newacronym{RQ}{RQ}{Rayleigh Quotient}
\newacronym{PF}{PF}{Perron-Frobenius}
\newacronym{RX}{RX}{Receiver}
\newacronym{BF}{BF}{Beamforming}
\newacronym{TX}{TX}{transitter}
\newacronym{TDL}{TDL}{Tap Delay Line}
\newacronym{QoS}{QoS}{Quality of Service}

\renewcommand{\smallskip}{\vspace{0.25cm}}

\newcommand{\abs}[1]{\big|{#1}\big|}

\newcommand{\tr}[1]{{\rm Tr}\left({#1}\right)}
\newcommand{\norm}[1]{||{#1}||}

\newcommand{\referencesrootdir}{./}
\newcommand{\myreferences}{\referencesrootdir/listofpublications.bib}

\begin{document}

\title{Full-Duplex Transmission Optimization for Bi-directional MIMO links with QoS Guarantees\\[-1ex]}

\author{
\IEEEauthorblockN{Hiroki Iimori$^\dagger$ and Giuseppe Abreu$^{\ddagger}$}
\IEEEauthorblockA{
$^\dagger$ Dep. of EEE, Ritsumeikan University\\ 1-1-1 Noji-higashi, Kusatsu, Shiga, 525-8577 Japan\\
{\tt [h.iimori,g.abreu]@gabreu.se.ritsumei.ac.jp}}
$^\ddagger$ Dep. of CS. \& EE., Jacobs University Bremen\\ Campus Ring 1, 28759, Bremen, Germany \\
{\tt g.abreu@jacobs-university.de}
\and
\IEEEauthorblockN{George C. Alexandropoulos}
\IEEEauthorblockA{Mathematical and Algorithmic Sciences Lab\\
Huawei Technologies France SASU\\
92100 Boulogne-Billancourt, France \\
{\tt george.alexandropoulos@huawei.com}}
}

\maketitle

\begin{abstract} 
We consider a bi-directional \ac{FD} \ac{MIMO} communication system in which nodes are capable of performing \ac{TX}-\ac{RX} digital precoding/combining and multi-tap analog cancellation, and have individual \ac{SINR} requirements.
We present an iterative algorithm for the TX powers minimization that includes closed-form expressions for the TX/RX digital beamformers at each algorithmic iteration step. 
Our representative simulation results demonstrate that the proposed algorithm can reduce residual \ac{SI} due to FD operation to below $-110$dB, which is the typical noise floor level for wireless communications.
In addition, our design outperforms relevant recent solutions proposed for 2-user MIMO systems (the so called MIMO X channel) in terms of both power efficiency and computational complexity.
\end{abstract}


\vspace{-0.5ex}
\section{Introduction}
\label{sect:intro}

Motivated by the exponentially increasing demand for higher information rate under limited wireless resources, and propelled by recent advances in \ac{RF} hardware, in-band \acf{FD} radio has emerged as a key technology for future wireless applications from \ac{5G} mobile communication systems to \ac{IoT} \cite{bharadia2013fullduplex ,Kim2015, Sabharwal2014SAiC}.

Practical communication in \ac{FD} mode requires dedicated solutions to mitigate the \acf{SI} caused by leakage of \ac{TX} signals into the \acf{RX} chain, due to the close proximity between \acf{TX} and \ac{RX} antennas \cite{Zhang2015CM,Choi2010MobiCom}.
Ironing out this fundamental issue of \ac{FD} technology is one of the main research topics in this field,  motivating various authors to contribute with several \ac{SI} cancellation techniques for \ac{FD} systems \cite{bharadia2013fullduplex, Everett2011Asilomar, Wang2015WCSP, Altieri2014SAiC}.

Thanks to the added \ac{DoF} afforded by multiple antennas, bi-directional \ac{FD} radio systems with high spectral efficiency can be designed exploiting \ac{MIMO} technology  \cite{vehkapera2013asymptotic, DaySP2012, jia2017signaling, Riihonen2013CISS}.
In particular, hybrid \ac{MIMO} \ac{SI} suppressing techniques combining analog and digital cancellers have been proposed for \ac{FD} radios \cite{GowdaTWC2018, MyListOfPapers:KorpiGlobecom2014, Sim2017CM, Cirik2013Asilomar} which proved very effective from a theoretical standpoint. 
From a practical implementation standpoint, however, it has been recently demonstrated in real-world experiments  that such \ac{MIMO} approaches are not devoid of its own technical challenges \cite{Jain2011MobiCom, korpi2016fullduplex, bharadia2013fullduplex}, one of which is the excessive cost incurred by the use of large numbers of antennas.

One approach to keep the hardware cost of hybrid \ac{MIMO} \ac{SI} suppressing techniques for \ac{FD} radios under control is to reduce the number of antennas while introducing temporal \acp{DoF} by means of \ac{TDL} processing in order to maintain the \ac{DoF} required to achieve the desired performance \cite{Alexandropoulos2017}.
In \cite{Alexandropoulos2017}, for instance, a joint hybrid \ac{TX}-\ac{RX} \ac{BF} design with limited hardware costs was proposed, in which the sum rate of a system with one  \ac{MIMO} \ac{FD} radio communicating with two  \ac{MIMO} \ac{HD} nodes was optimized.

In this paper, we contribute to the area of effective and feasible \ac{SI} canceller designs for \ac{MIMO} \ac{FD} radios as follows.
First, we combine the joint hybrid \ac{TX}-\ac{RX} approach of \cite{Alexandropoulos2017} with the analog cancellation technique referred to as \emph{multi-tap analog canceller} previously presented in \cite{Kolodziej2016TWC}.
The result is a new multi-tap \emph{hybrid} (analog and digital) \ac{TX}-\ac{RX} \ac{MIMO} \ac{FD} \ac{SI} cancellation scheme, in which the number of hardware components for analog cancellation becomes independent of the number of antennas.
Secondly, instead of maximizing the sum rate (which is of less practical interest), we formulate our problem to minimize the TX power while guaranteeing (when possible) prescribed \ac{QoS} targets defined in terms of maximum \acf{SINR}.
Thirdly and finally, we present a low-complexity solution to the latter problem in which
the \ac{TX} employs \ac{MRT} with powers optimized in closed-form via a \ac{PF} method, while the \ac{RX} maximizes the SINR by computing corresponding closed-form \ac{RX} \ac{BF} vectors from a \ac{RQ}, iteratively.
Our results show that our algorithm can outperform the similar methods previously proposed for 2-user MIMO systems in terms of both power efficiency and computational complexity.


\vspace{-0.6ex}
\section{System Model}
\label{System_Model}
\vspace{-0.2ex}

\begin{figure}[H]
\center
\includegraphics[width=\columnwidth]{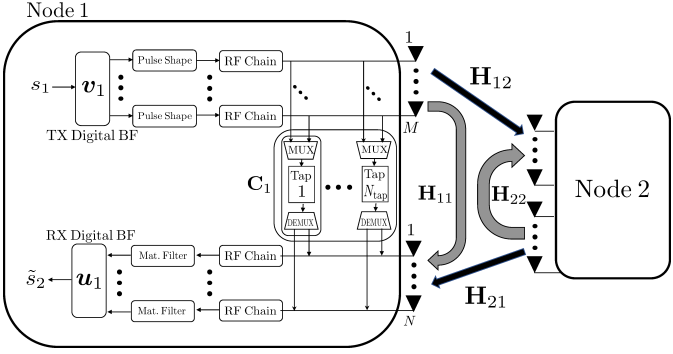}
\caption{System model of two-way full duplex MIMO with reduced hardware multi-tap analog cancellation.}
\label{fig:System_model}
\vspace{-1ex}
\end{figure}

Consider the two-way \ac{FD} \ac{MIMO} communication system illustrated in Figure \ref{fig:System_model}.
This system consists of two node in which each equipped with $M$ \ac{TX} and $N$ receive antennas.
Both nodes are assumed to \ac{TX} and receive simultaneously to/from one another in the same resource unit.

A generic $k$-th node, with $k\in\{1,2\}$, is assumed to employ the digital \ac{TX} precoding vector $\bm{v}_{k}\in\mathbb{C}^{M\times 1}$ and the digital \ac{RX} \ac{BF} vector $\bm{u}_{k}\in\mathbb{C}^{1\times N}$, as well as the multi-tap analog cancellation matrix $\bm{C}_{k}\in\mathbb{C}^{N\times M}$ \cite{Alexandropoulos2017, Kolodziej2016TWC}.
It is capable of performing TX-RX digital \ac{BF} and analog \ac{SI} cancellation with the aim at suppressing \ac{SI} and maximizing rate simultaneously. 
Finally, in order to model practical limitations, it is assumed that the \ac{TX}ted signal at the $k$-th node has a power upper bound, such that $\tr{\bm{v}_{k}\bm{v}^{\rm H}_{k}}= P_{k}\leq P_{\rm max}$.

Referring to Figure \ref{fig:System_model}, let $\bm{H}_{k\ell}\in\mathbb{C}^{N\times M}$ and $\bm{H}_{kk}\in\mathbb{C}^{N\times M}$ be the intended channel matrix between the two nodes and the \ac{SI} channel matrix at the $k$-th node, respectively, with $k\neq\ell\in\{1,2\}$.
It is also assumed throughout this paper that each node has full knowledge of the \ac{CSI} of both the communication links and their own \ac{SI} link.
Extension to imperfect \ac{CSI} knowledge is left for future work.

From all the above, the received signal at the $k$-th node after applying analog \ac{SI} cancellation can be written as
\begin{eqnarray}
\label{eq:received_k}
\bm{y}_{k} &=& \bm{H}_{\ell k}\bm{v}_{\ell}s_{\ell} + \left(\bm{H}_{kk} - \bm{C}_{k}\right)\bm{v}_{k}s_{k} + \bm{n}_{k}\\
&=& \underbrace{\bm{H}_{\ell k}\bm{v}_{\ell}s_{\ell}}_{\rm Intended} + \underbrace{\tilde{\bm{H}}_{kk}\bm{v}_{k}s_{k}}_{\rm SI} + \underbrace{\bm{n}_{k}}_{\rm Noise}\nonumber,
\end{eqnarray}
where the multi-tap analog cancellation matrix $\bm{C}_{k}$ consists of $N_{\rm tap}$ non-zero components and $MN - N_{\rm tap}$ zeros, $\bm{n}_{k}\sim \mathcal{CN}\left(0,\sigma^2\bm{I}_{N}\right)$ denotes the complex \ac{AWGN} vector under the assumption that $\bm{n}_{k}$ is independent from the \ac{TX}ted signal $s_{\ell}$, and $\tilde{\bm{H}}_{kk}\triangleq\bm{H}_{kk} - \bm{C}_{k}$ is the \ac{SI} channel matrix after performing the considered analog cancellation.

After digital down conversion and combining by the \ac{RX} \ac{BF} vector $\bm{u}_{k}$, the estimated signal $\tilde{s}_{\ell}$ corresponding to the intended signal $s_{\ell}$ at the $k$-th node can be expressed as
\begin{eqnarray}
\label{eq:estimated_k}
\tilde{s}_{\ell} &=& \bm{u}_{k}\bm{y}_{k}\\
&=&\bm{u}_{k}\bm{H}_{\ell k}\bm{v}_{\ell}s_{\ell} + \bm{u}_{k}\tilde{\bm{H}}_{kk}\bm{v}_{k}s_{k} + \bm{u}_{k}\bm{n}_{k}.\nonumber
\end{eqnarray}

Similarly, the received signal and symbol estimate at node $\ell\neq k$ after analog cancellation and \ac{RX} \ac{BF} are given, respectively, by
\begin{eqnarray}
\label{eq:received_l}
\bm{y}_{\ell} &=& \bm{H}_{k\ell}\bm{v}_{k}s_{k} + \tilde{\bm{H}}_{\ell\ell}\bm{v}_{\ell}s_{\ell} + \bm{n}_{\ell},\\
\label{eq:estimated_l}
\tilde{s}_{k} &=& \bm{u}_{\ell}\bm{y}_{\ell}\nonumber\\
&=&\bm{u}_{\ell}\bm{H}_{k\ell}\bm{v}_{k}s_{k} + \bm{u}_{\ell}\tilde{\bm{H}}_{\ell\ell}\bm{v}_{\ell}s_{\ell} + \bm{u}_{\ell}\bm{n}_{\ell},
\end{eqnarray}
where $\bm{n}_{\ell}\sim \mathcal{CN}\left(0,\sigma^2\bm{I}_{N}\right)$ is the AWGN vector that is assumed independent from the \ac{TX}ted symbol $s_{k}$. 

Assuming that unit power information signals $s_{k}$ and $s_{\ell}$ are used, the average SINR estimates at the two nodes in Figure \ref{fig:System_model} can be, respectively, written as
\begin{equation}
\label{eq:SINR}
\gamma_{k} = \frac{\abs{\bm{u}_{k}\bm{H}_{\ell k}\bm{v}_{\ell}}^2}{\abs{\bm{u}_{k}\tilde{\bm{H}}_{kk}\bm{v}_{k}}^2 + \sigma^2} \quad \text{and}\quad 
\gamma_{\ell} = \frac{\abs{\bm{u}_{\ell}\bm{H}_{k\ell}\bm{v}_{k}}^2}{\abs{\bm{u}_{\ell}\tilde{\bm{H}}_{\ell\ell}\bm{v}_{\ell}}^2 + \sigma^2},
\end{equation}
where we assume that the channel matrices in equation \eqref{eq:SINR} are constant for a number of signal transmissions and the \ac{RX} combining vector $\bm{u}_{k},\forall k$ has a unit norm, i.e., $\norm{\bm{u}_{k}}^2=1$.

\vspace{-1ex}
\section{QoS-Guaranteed Transmissions}
\label{sect:OptimalTS}
\vspace{-1ex}

Signal processing techniques for the joint TX-RX linear precoding/combining and adaptive \ac{TX} power allocation with the aim of maximizing data rate while suppressing the residual \ac{SI} power level have been proposed in the past \cite{Zheng2015TSP, IimoriSPAWC2018} demonstrating the feasibility of two-way \ac{FD} \ac{MIMO} systems.

Maximizing data rate is, however, not typically required by actual users, which instead tend to perceive the
quality of a communication system by comparing it to a given level of expectation dictated by the intended application.
We therefore consider instead the TX-RX beamformer optimization problem aiming at minimizing the individual \ac{TX} powers while satisfying individual target SINR requirements:
\begin{subequations}
\label{MMES_OP_Problem}
\begin{eqnarray}
\label{OP1}
\min_{\bm{v}_{k},\bm{v}_{\ell}}&& \sum^{2}_{k=1}\norm{\bm{v}_{k}}^2\\
\label{ConstraintP}
{\rm s.t.}&& \gamma_{k} \geq \Gamma_{k}\:\: \forall k,
\end{eqnarray}
\end{subequations}
where $\Gamma_{k}$ is the target SINR for the $k$-th node.

\vspace{-0.5ex}
\subsection{\ac{TX} Power Minimization with SINR Constraints}
\label{sect:minpower}
\vspace{-0.5ex}

Let us define the normalized precoding vector $\bar{\bm{v}}_{k}\triangleq\frac{\bm{v}_{k}}{\norm{\bm{v}_{k}}}$ and the \ac{TX} power $P_{k}=\norm{\bm{v}_{k}}^2$ such that the optimization problem in \eqref{MMES_OP_Problem} can be rewritten as
\begin{subequations}
\label{OP2}
\vspace{-0.5ex}
\begin{eqnarray}
\min_{P_{1},P_{2}}&& \sum^{2}_{k=1}P_{k}\\
\label{ConstraintP}
{\rm s.t.}&& \gamma_{k} \geq \Gamma_{k} \:\: \forall k.
\end{eqnarray}
\end{subequations}

The optimization problem described by equation \eqref{OP2} is well-known to be non convex due to the SINR constraints \cite{Malla2015CSIT}, although approximate solutions can be obtained for it with basis on convex optimization algorithms, such as interior point methods if the constraint can be convexified \cite{Boyd2004}.
In addition to the losses due to convex relaxation, such solutions tend also to be computationally demanding.
Therefore, we propose instead a low complexity alternating minimization method based on closed-form expressions of the optimal \ac{TX} powers $P_{1}$ and $P_{2}$.

In order to obtain the desired closed-form expressions for $P_{1}$ and $P_{2}$, notice that from equation \eqref{eq:SINR} and \eqref{OP2} we readily obtain
\begin{subequations}
\label{eqn:ineqPopt}
\begin{eqnarray}
&P_{2}\abs{\bm{u}_{1}\bm{H}_{21}\bar{\bm{v}}_{2}}^2 \geq \Gamma_{1}\Big({P_{1}\abs{\bm{u}_{1}\tilde{\bm{H}}_{11}\bar{\bm{v}}_{1}}^2 + \sigma^2}\Big),\\
&P_{1}\abs{\bm{u}_{2}\bm{H}_{12}\bar{\bm{v}}_{1}}^2 \geq \Gamma_{2}\left({P_{2}\abs{\bm{u}_{2}\tilde{\bm{H}}_{22}\bar{\bm{v}}_{2}}^2 + \sigma^2}\right).
\end{eqnarray}
\end{subequations}
The latter inequalities can be re-expressed in matrix form as
\begin{eqnarray}
\label{eqn:ineqPoptMat}
\left({\bf{I}}- \bm{\Gamma}\bm{M}\right)\bm{p}\geq\sigma^2 \bm{\Gamma} \bm{m},
\end{eqnarray}
where we define the \ac{TX} power vector $\bm{p} \triangleq \left[P_{1},P_{2}\right]^{\rm T}$ and the auxiliary matrices $\bm{\Gamma}$, $\bm{M}$ and $\bm{m}$ respectively  by

\begin{subequations}
\begin{eqnarray}
\bm{\Gamma} &=& 
\begin{bmatrix}
0 & \Gamma_{2} \\
\Gamma_{1} & 0
\end{bmatrix},\\
\bm{M} &=& 
\begin{bmatrix}
\frac{|\bm{u}_{1}\tilde{\bm{H}}_{11}\bar{\bm{v}}_{1}|^2}{|\bm{u}_{1}\bm{H}_{21}\bar{\bm{v}}_{2}|^2} & 0\\
0 & \frac{|\bm{u}_{2}\tilde{\bm{H}}_{22}\bar{\bm{v}}_{2}|^2}{|\bm{u}_{2}\bm{H}_{12}\bar{\bm{v}}_{1}|^2}
\end{bmatrix},\\
\bm{m} &=& 
\begin{bmatrix}
\abs{\bm{u}_{1}\bm{H}_{21}\bar{\bm{v}}_{2}}^{-2} &
\abs{\bm{u}_{2}\bm{H}_{12}\bar{\bm{v}}_{1}}^{-2}
\end{bmatrix}^\text{T}.
\end{eqnarray}
\end{subequations}

Taking advantage of the \ac{PF} theorem \cite{S-U-Pillai2005SPM} and the fact that $\bm{\Gamma}\bm{M}$ is a non negative matrix, the optimal \ac{TX} power vector can be computed in closed form as
\begin{eqnarray}
\label{eqn:popt}
\bm{p}^{*}=\sigma^2\left({\bf{I}}- \bm{\Gamma}\bm{M}\right)^{-1} \bm{\Gamma} \bm{m}.
\end{eqnarray}

\subsection{Optimal \ac{BF} Design for SINR Maximization}
\label{sect:beamdesign}

With possession of a closed-form optimal solution to the \ac{TX} power vector $\bm{p}$ as per equation \eqref{eqn:popt}, as well as a given analog cancellation matrix $\bm{C}_{k}$ obtained for example as discussed in \cite{Alexandropoulos2017, IimoriSPAWC2018}, we seek optimal \ac{BF} designs for $\bm{v}_{k}$ and $\bm{u}_{k}\: \forall k$, such that the average SINR at each node is maximized, while minimizing the effect of the \ac{SI}.
Taking into account the fact that the role of TX-RX beamformers is to minimize the effect of \ac{SI} while maximizing the downlink rate, we consider the \ac{MRT} TX beamformer with perfect \ac{CSI} known at the nodes, such that the instantaneous \ac{SINR} at each node is maximized under the assumption that the \ac{SI} power level can be significantly reduced after processing by the proposed optimal \ac{RX} combiner.

\vspace{1ex}
\subsubsection[]{Design of \ac{RX} Combiner $\bm{u}_{k}\forall k$}\quad\\[-2ex]

The role of the \ac{RX} combining vector $\bm{u}_{k}$ at the $k$-th node is to maximize the power of the signal from the $\ell$-th node, while supressing the interference-plus-noise signal.
In other words, the \ac{RX} \ac{BF} vector $\bm{u}_{k}$ must be designed so as to maximize the ratio between the power of the intended signal and that of interference-plus-noise term of equation \eqref{eq:estimated_k}, which can be mathematically expressed as
\begin{eqnarray}
\label{eq:gk_OP}
\max_{\bm{u}_{k} \atop \norm{\bm{u}_{k}}^2=1} \frac{\bm{u}_{k}\overbrace{\bm{H}_{\ell k}\bm{v}_{\ell}\bm{v}^{\rm H}_{\ell}\bm{H}^{\rm H}_{\ell k}}^{\triangleq \bm{Q}_{\bm{u}_{k}}}\bm{u}^{\rm H}_{k}}{\bm{u}_{k}\underbrace{\left(\tilde{\bm{H}}_{k k}\bm{v}_{k}\bm{v}^{\rm H}_{k}\tilde{\bm{H}}^{\rm H}_{k k} + \sigma^2{\bf{I}}\right)}_{^{\triangleq \bm{W}_{\bm{u}_{k}}}}\bm{u}^{\rm H}_{k}},
\end{eqnarray}
which holds a generalized \ac{RQ} structure, such that the optimal solution to $\bm{u}_{k}$ is obtained by \cite{Prieto2003ICASSP}
\begin{eqnarray}
\label{eq:ukopt}
\bm{u}^{*}_{k} = {\rm eigv}_{\rm max}\left(\bm{W}^{-1}_{\bm{u}_{k}}\bm{Q}_{\bm{u}_{k}}\right)^{\rm H}.
\end{eqnarray}

\begin{figure}[H]
\center
\vspace{-1ex}
\includegraphics[width=\columnwidth]{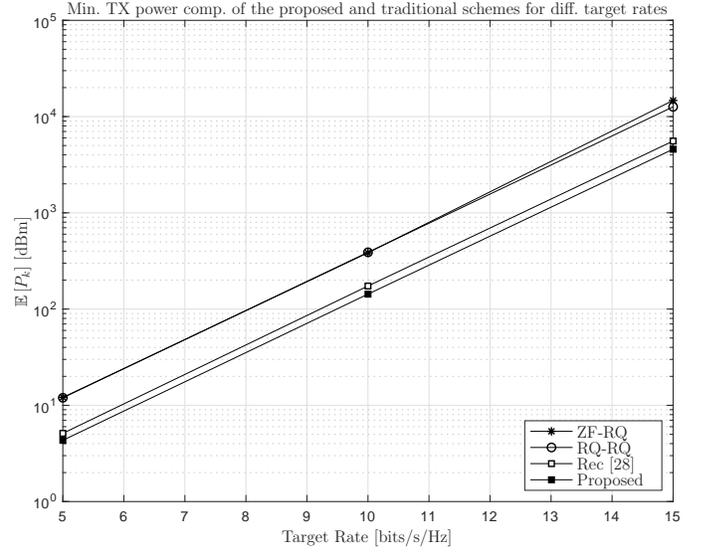}
\vspace{-3.5ex}
\caption{Proposed and conventional preceding methods TX power comparison for different target rates.}
\label{fig:TransPower}
\vspace{-3ex}
\end{figure}
\begin{algorithm}[h]
\caption[]{Alt. \ac{TX} Power Min. with SINR Guarantees.}
\label{alg:main}
\begin{algorithmic}[1]
\State \textbf{Input}:$P_{k},\bm{H}_{kk},\bm{H}_{\ell k}$, \:$\bm{C}_{k} \forall k$ given by \cite{Alexandropoulos2017}.
\State Set $P_{k}=P_{\rm max} \forall k\in\{1,2\}$ and make arbitrary unit-norm vectors as initial \ac{RX} \ac{BF} vectors $\bm{u}_{k} \forall k$.
\Repeat
\State Compute $\bar{\bm{v}}_{k}\forall k$ from equation \eqref{eqn:MRTTX}.
\State Compute $\bm{u}^{*}_{k}\forall k$ from equation \eqref{eq:ukopt}.
\State Compute $\bm{p}^{*}$ from equation \eqref{eqn:popt}.
\Until{{\rm\bf convergence or reach maximum iterations.}}
\end{algorithmic}
\end{algorithm}
\setlength{\textfloatsep}{-2pt}

\vspace{-2ex}
\subsubsection{Design of TX Precoder $\bm{v}_{k}\forall k$}\quad\\[-2.3ex]

Assuming that the strong \ac{SI} caused by the leakage of own \ac{TX} signals due to the close proximity of \ac{TX} and \ac{RX} antennas can be sufficiently suppressed by the \ac{RX} combining vector $\bm{u}_{k}$, the role of the TX precoder $\bm{v}_{k}$ is only to direct the \ac{TX} beams so as to maximize the downlink rate.
For this purpose, it suffices to apply a simple \ac{MRT} TX precoder, namely
\vspace{-1ex}
\begin{eqnarray}
\label{eqn:MRTTX}
\vspace{-1ex}
\bar{\bm{v}}_{k} = \frac{\bm{H}^{\rm H}_{kl}\bm{u}^{\rm H}_{\ell}}{\left|\left|\bm{H}_{kl}\bm{u}_{\ell}\right|\right|}.
\vspace{-1ex}
\end{eqnarray}

Taking into account all the steps described in this section, the proposed algorithm for the optimization of \ac{TX} powers, as well as \ac{TX} and \ac{RX} \ac{BF} vectors can be compactly described by the pseudo-code offered in Algorithm \ref{alg:main}.

\vspace{-1ex}
\section{Simulation results}
\label{SimulationResults}
\vspace{-0.5ex}

In this section, we evaluate the proposed iterative algorithm in terms of consumed \ac{TX} power for different required SINR constraints via software simulation.
The downlink communication channels $\bm{H}_{12}$ and $\bm{H}_{21}$ are assumed to be block Rayleigh fading with $110$dB path loss, while the \ac{SI} channels $\bm{H}_{11},\bm{H}_{22}$ are assumed to be block Ricean with path loss of $40$dB and $K$-factor $35$dB \cite{Alexandropoulos2017,duarte2012experiment}.

Each node is assumed to be equipped with $4$ \ac{TX} and receive antennas, i$.$e$.$, $M=N=4$, with a noise floor of $-110$dBm, and the number of analog cancellation taps $N_{\rm tap}$ is set to $8$, which corresponds to $50$\% reduction in the number of elements in the analog cancellation matrix $\bm{C}_{k}$ compared to \cite{bharadia2013fullduplex, Everett2011Asilomar}.
\begin{figure}[H]
\center
\vspace{1ex}
\includegraphics[width=\columnwidth]{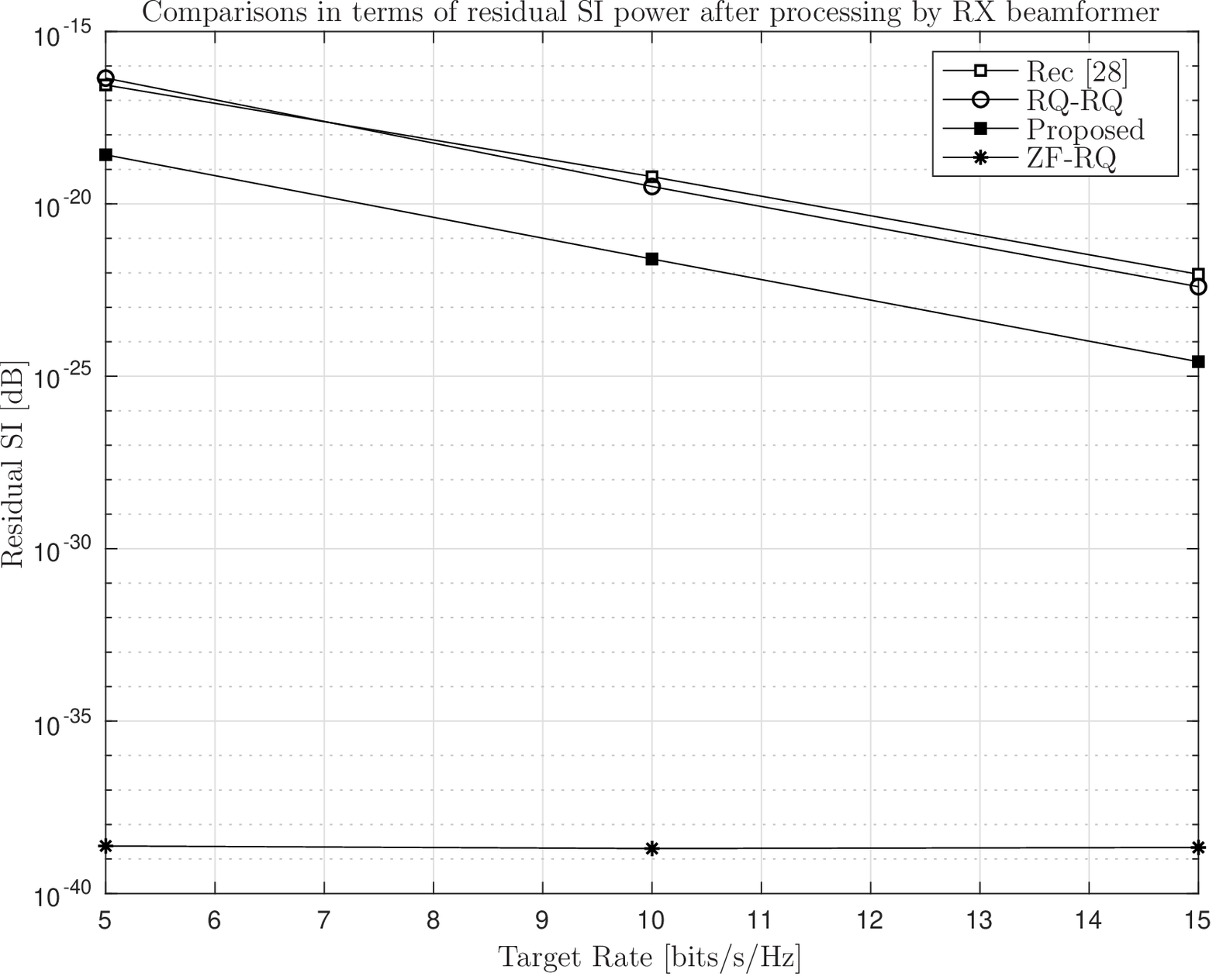}
\vspace{-4ex}
\caption{Residual SI power comparisons of the proposed and conventional methods for different target rates after \ac{RX} \ac{BF}.}
\label{fig:ResidualSI}
\vspace{1ex}
\includegraphics[width=\columnwidth]{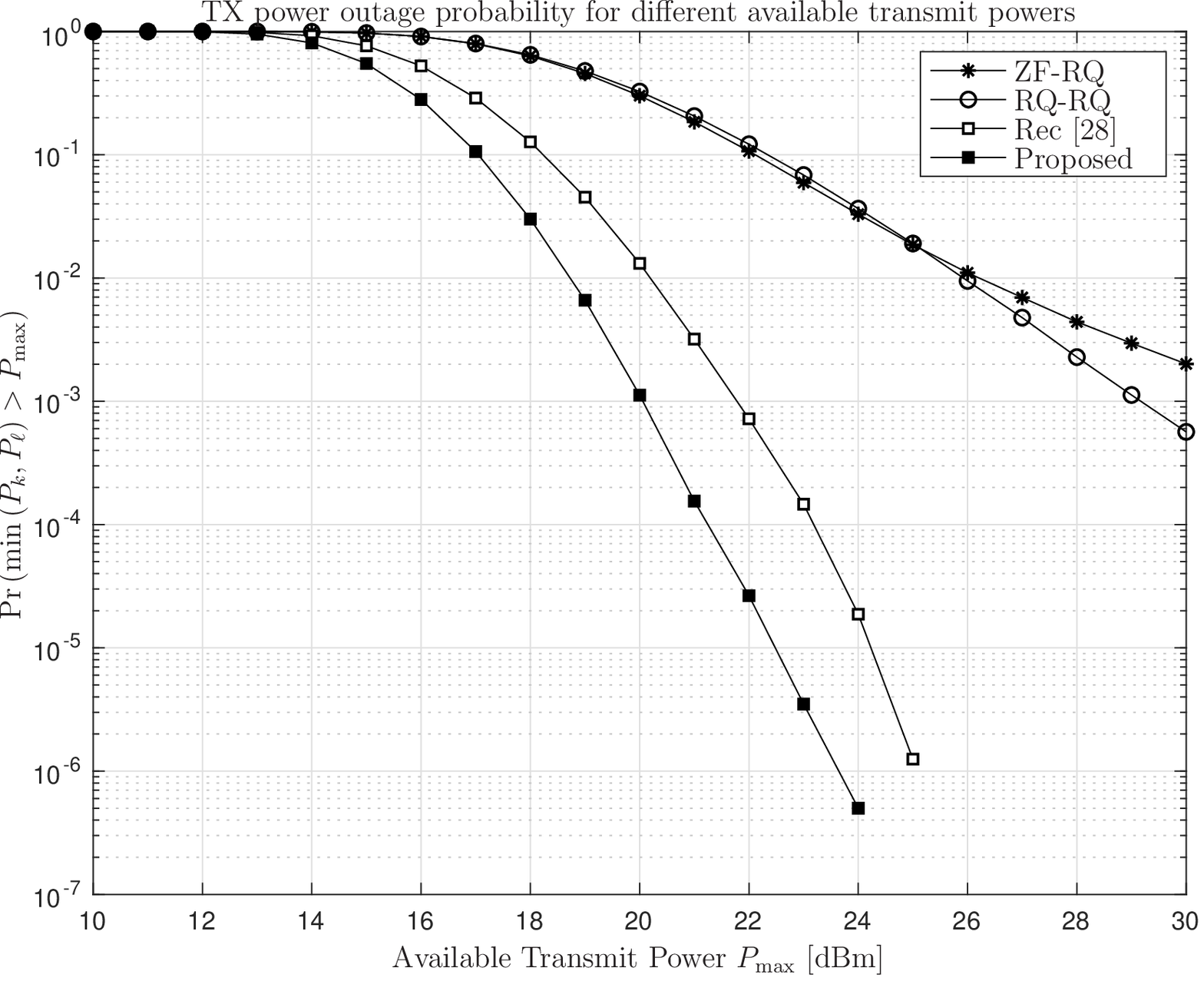}
\vspace{-4ex}
\caption{\ac{TX} power outage probability for different available \ac{TX} powers with the fixed target rate $R_{k}=R_{\ell}=8$ [bps/Hz].}
\label{fig:OutPower}
\vspace{-2ex}
\end{figure}
\noindent For the modeling practical situations, the multi-tap analog canceller is assumed to be subjected to amplitude imperfection uniformly distributed between $-0.01$dB and $0.01$dB and phase noise uniformly distributed between $-0.065^{\circ}$ and $0.065^{\circ}$ \cite{Alexandropoulos2017, Kolodziej2016TWC}. 

In all figures that follow, we compare the proposed \ac{TX} power minimization method in Algorithm \ref{alg:main} with $100$ maximum iterations against the conventional \ac{ZF} TX precoder, in which the proposed \ac{PF} power optimization is applied. 
In addition, by noticing that the considered bi-directional FD MIMO corresponds to a special form of the MIMO X channel, we deploy relevant algorithms \cite{GeorgeSP2013, Alexandropoulos2016_CB, OmidTWC2018} targeting at TX-RX \ac{BF} design yielding sum rate maximization. 
Particularly, our considered system is a MIMO X channel having $\bm{H}_{21}$ and $\bm{H}_{12}$ as the intended channels and $\tilde{\bm{H}}_{11}$ and $\tilde{\bm{H}}_{22}$ as the interference channels, having possibly larger powers than the intended ones.

\begin{table}[t!]
\caption{Run time comparisons for different methods.}
\vspace{-3ex}
\begin{center}
\begin{tabular}{|c|c|c|c|c|} \hline
  Methods  &  ZF & RQ-RQ & Rec & Proposed \\ \hline 
  Average run time [s] & 0.0025 & 0.0028 & 0.1309 & 0.0022 \\\hline
\end{tabular}
\vspace{-1ex}
\label{Tab:Runtime}
\end{center}
\end{table}

First, average \ac{TX} power comparisons of the proposed algorithm for different target rates $\log_{2}(1+{\Gamma_{k}}) \forall k$ is shown in Figure \ref{fig:TransPower}, where \ac{ZF}-\ac{RQ}, \ac{RQ}-\ac{RQ} \cite{OmidTWC2018} and the Reconfigurable sum rate maximization algorithm \cite{GeorgeSP2013} are employed as a benchmark.
In order to fairly compare those algorithms, we adopt an alternating recalculation between \ac{TX}-\ac{RX} \ac{BF} for each algorithms until convergence or maximum number of iterations reached.
It is shown in Figure \ref{fig:TransPower} that the proposed method can decrease the \ac{TX} power by about $-4.5$dB compared to the conventional \ac{ZF}-\ac{RQ}, \ac{RQ}-\ac{RQ} methods and about $-0.8$dB compared against the Reconfigurable method.

Secondly, Figure \ref{fig:ResidualSI} outlines that the interference cancellation performance in terms of residual \ac{SI} power levels after processing by the \ac{RX} \ac{BF} are compared for the different \ac{TX}-\ac{RX} \ac{BF} schemes.
From Figure \ref{fig:TransPower} and \ref{fig:ResidualSI}, one can notice that although the \ac{ZF} method can perfectly suppress the effect of \ac{SI} at the \ac{RX} baseband, the proposed method can outperform the other schemes due to the fact that not only the residual \ac{SI} level of the proposed method is suppressed below the noise floor level but also it aims at maximizing the data rate performance.
In other words, the other methods devote too much available \ac{DoF}s to suppressing \ac{SI} power level at the \ac{RX} baseband.

Thirdly, the \ac{TX} power outage probability of the proposed method for different available \ac{TX} powers $P_{\rm max}$ with target data rate fixed at $R_{k}=R_{\ell}=8$ [bps/Hz] is compared with the outage performance of the other conventional methods in Figure \ref{fig:OutPower}, where we define the \ac{TX} power outage probability as ${\rm Pr}\left({\rm min}\left(P_{k}, P_{\ell}\right)>P_\mathrm{max}\right)$.
Lastly, the average run time comparisons until the convergence for each different algorithms are depicted in TABLE \ref{Tab:Runtime}, where we take an average from $500$ channel realizations.
From Figure \ref{fig:TransPower}, \ref{fig:ResidualSI} and \ref{fig:OutPower} and Table \ref{Tab:Runtime}, it can be observed that the proposed method can has much fast convergence rate compared with the Reconfigurable method and outperform the conventional \ac{ZF} and \ac{RQ} \ac{TX} \ac{BF} methods in terms of the \ac{TX} power outage probability performance.

\vspace{-0.4ex}
\section{Conclusion}
\label{Conclusion}
\vspace{-0.6ex}
In this paper, we considered bi-directional \ac{FD} \ac{MIMO} communications systems with limited number of analog canceller taps and designed TX-RX \ac{BF} vectors with the goal to minimize \ac{TX} power under \ac{SINR} constraints.

The proposed \ac{TX} power minimization \ac{BF} design was investigated in terms of system performance and complexity, and the \ac{PF} \ac{TX} power minimization approach was jointly offered with the proposed beamformers.
Simulation results demonstrate the capability of our proposed algorithm to suppress the \ac{SI} level to below $-110$dB which is the typical noise floor for wireless communications, while maximizing the downlink rate, and consequently, it minimizes the average \ac{TX} power for different target data rate.

\vspace{-0.4ex}
\section{Acknowledgement}
\vspace{-0.6ex}

Parts of this work were supported by JSPS KAKENHI, Grant Number JP17K06448, Denkitsushin Hukyu Foundation, and EU Project HIGHTS with grant number 636537.

\bibliographystyle{IEEEtran}
\bibliography{IEEEabrv,\myreferences,MyListOfPapers}

\begin{thebibliography}{10}
\providecommand{\url}[1]{#1}
\csname url@samestyle\endcsname
\providecommand{\newblock}{\relax}
\providecommand{\bibinfo}[2]{#2}
\providecommand{\BIBentrySTDinterwordspacing}{\spaceskip=0pt\relax}
\providecommand{\BIBentryALTinterwordstretchfactor}{4}
\providecommand{\BIBentryALTinterwordspacing}{\spaceskip=\fontdimen2\font plus
\BIBentryALTinterwordstretchfactor\fontdimen3\font minus
  \fontdimen4\font\relax}
\providecommand{\BIBforeignlanguage}[2]{{%
\expandafter\ifx\csname l@#1\endcsname\relax
\typeout{** WARNING: IEEEtran.bst: No hyphenation pattern has been}%
\typeout{** loaded for the language `#1'. Using the pattern for}%
\typeout{** the default language instead.}%
\else
\language=\csname l@#1\endcsname
\fi
#2}}
\providecommand{\BIBdecl}{\relax}
\BIBdecl

\bibitem{bharadia2013fullduplex}
D.~Bharadia, E.~McMilin, and S.~Katti, ``{Full duplex radios},'' in
  \emph{{Proc. ACM SIGCOMM.}}, New York, NY, USA, 2013, pp. 375--386.

\bibitem{Kim2015}
D.~Kim, H.~Lee, and D.~Hong, ``A survey of in-band full-duplex transmission:
  From the perspective of {PHY} and {MAC} layers,'' \emph{{IEEE Commun. Surv.
  Tuts.}}, vol.~17, no.~4, pp. 2017--2046, Feb. 2015.

\bibitem{Sabharwal2014SAiC}
A.~Sabharwal, P.~Schniter, D.~Guo, D.~W. Bliss, S.~Rangarajan, and R.~Wichman,
  ``In-band full-duplex wireless: Challenges and opportunities,'' \emph{{IEEE
  J. Sel. Areas Commun.}}, vol.~32, no.~9, pp. 1637--1652, June 2014.

\bibitem{Zhang2015CM}
Z.~Zhang, X.~Chai, K.~Long, A.~V. Vasilakos, and L.~Hanzo, ``Full duplex
  techniques for {5G} networks: {Self}-interference cancellation, protocol
  design, and relay selection,'' \emph{{IEEE Commun. Mag.}}, vol.~53, no.~5,
  pp. 128--137, May 2015.

\bibitem{Choi2010MobiCom}
J.~I. Choi, M.~Jain, K.~Srinivasan, P.~Levis, and S.~Katti, ``Achieving single
  channel, full duplex wireless communication,'' in \emph{{Proc. MOBICOM}}, New
  York, NY, USA, Sep. 20 - 24 2010, pp. 1--12.

\bibitem{Everett2011Asilomar}
E.~Everett, M.~Duarte, C.~Dick, and A.~Sabharwal, ``Empowering full-duplex
  wireless communication by exploiting directional diversity,'' in
  \emph{{P}roc. Asilomar {CSSC}}, Nov. 2011, pp. 2002--2006.

\bibitem{Wang2015WCSP}
Y.~Wang, Q.~Jiang, Z.~Chen, and B.~Xia, ``Outage probability of two-way
  full-duplex amplify-forward relay systems with asymmetric traffic
  requirements,'' in \emph{{Proc. IEEE WCSP}}, Nanjing, China, 15-17 Oct. 2015,
  pp. 1--5.

\bibitem{Altieri2014SAiC}
A.~Altieri, L.~R. Vega, P.~Piantanida, and C.~G. Galarza, ``On the outage
  probability of the full-duplex interference-limited relay channel,''
  \emph{{IEEE J. Sel. Areas Commun.}}, vol.~32, no.~9, pp. 1--13, Sep. 2014.

\bibitem{vehkapera2013asymptotic}
M.~Vehkaper\"{a}, T.~Riihonen, and R.~Wichman, ``Asymptotic analysis of
  full-duplex bidirectional {MIMO} link with transmitter noise,'' in
  \emph{{Proc. IEEE PIMRC}}, London, UK, Sep. 2013, pp. 1265--1270.

\bibitem{DaySP2012}
B.~Day, A.~Margetts, D.~Bliss, and P.~Schniter, ``Full-duplex bidirectional
  {MIMO}: Achievable rates under limited dynamic range,'' \emph{{IEEE Trans.
  Signal Process.}}, vol.~60, no.~7, pp. 3702--3713, Jul. 2012.

\bibitem{jia2017signaling}
S.~Jia and B.~Aazhang, ``Signaling design of two-way {MIMO} full-duplex
  channel: Optimality under imperfect transmit front-end chain,'' \emph{{IEEE
  Trans. Wireless Commun.}}, vol.~16, no.~3, pp. 1619--1632, March 2017.

\bibitem{Riihonen2013CISS}
T.~Riihonen, M.~Vehkaper\"a, and R.~Wichman, ``Large-system analysis of rate
  regions in bidirectional full-duplex {MIMO} link: Suppression versus
  cancellation,'' in \emph{{Proc. IEEE CISS}}, Baltimore, USA, Mar. 2013, pp.
  1--6.

\bibitem{GowdaTWC2018}
N.~M. Gowda and A.~Sabharwal, ``Jointnull: Combining reconfigurable analog
  cancellation with transmit beamforming for large-antenna full-duplex
  wireless,'' \emph{{IEEE Trans. Wireless Commun.}}, vol.~17, no.~3, pp.
  2094--2108, Mar. 2018.

\bibitem{MyListOfPapers:KorpiGlobecom2014}
D.~Korpi, L.~Anttila, and M.~Valkama, ``Reference receiver based digital
  self-interference cancellation in mimo full-duplex transceivers,'' in
  \emph{{Proc. IEEE GLOBECOM}}, Austin, USA, Dec. 2014.

\bibitem{Sim2017CM}
M.~S. Sim, M.~Chung, D.~Kim, J.~Chung, D.~K. Kim, and C.-B. Chae, ``Nonlinear
  self-interference cancellation for full-duplex radios: From link-level and
  system-level performance perspectives,'' \emph{{IEEE Commun. Mag.}}, vol.~55,
  no.~9, Jun. 2017.

\bibitem{Cirik2013Asilomar}
A.~C. Cirik, R.~Wang, and Y.~Hua, ``Weighted-sum-rate maximization for
  bi-directional full-duplex {MIMO} systems,'' in \emph{{P}roc. Asilomar
  {CSSC}}, Nov. 2013.

\bibitem{Jain2011MobiCom}
M.~Jain, J.~I. Choi, T.~M. Kim, D.~Bharadia, S.~Seth, K.~Srinivasan, P.~Levis,
  S.~Katti, and P.~Sinha, ``Practical, real-time, full duplex wireless,'' in
  \emph{{Proc. ICMCN}}, New York, USA, Sep. 2011, pp. 301--312.

\bibitem{korpi2016fullduplex}
D.~Korpi, J.~Tamminen, M.~Turunen, T.~Huusari, Y.~S. Choi, L.~Anttila,
  S.~Talwar, and M.~Valkama, ``Full-duplex mobile device: {P}ushing the
  limits,'' \emph{{IEEE Commun. Mag.}}, vol.~54, no.~9, pp. 80--87, Sep. 2016.

\bibitem{Alexandropoulos2017}
G.~C. Alexandropoulos and M.~Duarte, ``Joint design of multi-tap analog
  cancellation and digital beamforming for reduced complexity full duplex
  {MIMO} systems,'' in \emph{Proc. IEEE ICC}, Paris, France, 21--25 May 2017,
  pp. 1--7.

\bibitem{Kolodziej2016TWC}
K.~E. Kolodziej, J.~G. McMichael, and B.~T. Perry, ``Multitap {RF} canceller
  for in-band full-duplex wireless communications,'' \emph{{IEEE Trans.
  Wireless Commun.}}, vol.~15, no.~6, pp. 4321--4334, Jun. 2016.

\bibitem{Zheng2015TSP}
G.~Zheng, ``Joint beamforming optimization and power control for full-duplex
  {MIMO} two-way relay channel,'' \emph{{IEEE Trans. Signal Process.}},
  vol.~63, no.~3, pp. 555--566, Feb. 2015.

\bibitem{IimoriSPAWC2018}
H.~Iimori and G.~Abreu, ``Two-way full-duplex {MIMO} with hybrid {TX-RX} {MSE}
  minimization and interference cancellation,'' in \emph{{Proc. IEEE SPAWC}},
  Kalamata, Greece, Jun. 2018, pp. 1--6.

\bibitem{Malla2015CSIT}
S.~Malla and G.~Abreu, ``Transmission strategies under imperfect instantaneous
  {CSIT},'' in \emph{{Proc. IEEE WCNC}}, New Orleans, USA, Mar. 2015, pp. 1--6.

\bibitem{Boyd2004}
S.~P. Boyd and L.~Vandenberghe, \emph{Convex Optimization}.\hskip 1em plus
  0.5em minus 0.4em\relax Cambridge University Press, 2004.

\bibitem{S-U-Pillai2005SPM}
S.~U. Pillai, T.~Suel, and S.~Cha, ``The {Perron-Frobenius} theorem--{Some of
  its applications},'' \emph{IEEE Signal Processing Mag.}, vol.~22, no.~2, pp.
  62--75, March 2005.

\bibitem{Prieto2003ICASSP}
R.~Prieto, ``A general solution to the maximization of the multidimensional
  generalized {Rayleigh} quotient used in linear discriminant analysis for
  signal classification,'' in \emph{{Proc. IEEE ICASSP}}, vol.~6, Hong Kong,
  China, Apr. 2003, pp. 1--4.

\bibitem{duarte2012experiment}
M.~Duarte, C.~Dick, and A.~Sabharwal, ``Experiment-driven characterization of
  full-duplex wireless systems,'' \emph{{IEEE Trans. Wireless Commun.}},
  vol.~11, no.~12, pp. 4296--4307, Dec. 2012.

\bibitem{GeorgeSP2013}
G.~C. Alexandropoulos and C.~B. Papadias, ``A reconfigurable iterative
  algorithm for the {$K$}-user {MIMO} interference channel,'' \emph{Signal
  Process.}, vol.~93, no.~12, pp. 3353--3362, Jun. 2013.

\bibitem{Alexandropoulos2016_CB}
G.~C. Alexandropoulos, P.~Ferrand, J.-M. Gorce, and C.~B. Papadias, ``Advanced
  coordinated beamforming for the downlink of future {LTE} cellular networks,''
  \emph{{IEEE Commun. Mag.}}, vol.~54, no.~7, pp. 54--60, Jul. 2016.

\bibitem{OmidTWC2018}
O.~Taghizadeh, A.~C. Cirik, and R.~Mathar, ``Hardware impairments aware
  transceiver design for full-duplex amplify-and-forward {MIMO} relaying,''
  \emph{{IEEE Trans. Wireless Commun.}}, vol.~17, no.~3, pp. 1644--1659, Dec.
  2018.

\end{thebibliography}
\end{document}